\PassOptionsToPackage{table,xcdraw}{xcolor}

\documentclass[sigconf]{acmart}

\setcopyright{acmcopyright}
\copyrightyear{2024}
\acmYear{2024}
\setcopyright{rightsretained}
\acmConference[ASSETS '24]{The 26th International ACM SIGACCESS Conference on Computers and Accessibility}{October 27--30, 2024}{St. John's, NL, Canada}
\acmBooktitle{The 26th International ACM SIGACCESS Conference on Computers and Accessibility (ASSETS '24), October 27--30, 2024, St. John's, NL, Canada}
\acmDOI{10.1145/3663548.3688498}
\acmISBN{979-8-4007-0677-6/24/10}

\usepackage{subfigure}

\usepackage{graphicx}
\usepackage{multirow}
\usepackage{pifont}
\usepackage{lscape}
\usepackage{tabularray}

\setcopyright{none}
\begin{document}

\title[An AI Guide to Enhance Accessibility of Social Virtual Reality for Blind People]{An AI Guide to Enhance Accessibility of Social Virtual Reality for Blind People}

\author{Jazmin Collins}
\email{jc2884@cornell.edu}
\affiliation{%
  \institution{Cornell University}
  \city{Ithaca}
  \state{New York}
  \country{USA}
}

\author{Kaylah Myranda Nicholson}
\email{kmn67@cornell.edu}
\affiliation{%
  \institution{Cornell University}
  \city{Ithaca}
  \state{New York}
  \country{USA}
}

\author{Yusuf Khadir}
\email{yk647@cornell.edu}
\affiliation{%
  \institution{Cornell University}
  \city{Ithaca}
  \state{New York}
  \country{USA}
}

\author{Andrea Stevenson Won}
\email{asw248@cornell.edu}
\affiliation{%
  \institution{Cornell University}
  \city{Ithaca}
  \state{New York}
  \country{USA}
}

\author{Shiri Azenkot}
\email{shiri.azenkot@cornell.edu}
\affiliation{%
  \institution{Cornell Tech}
  \city{New York}
  \state{New York}
  \country{USA}
}

\renewcommand{\shortauthors}{Collins, et al.}

\begin{abstract}
The rapid growth of virtual reality (VR) has led to increased use of social VR platforms for interaction. However, these platforms lack adequate features to support blind and low vision (BLV) users, posing significant challenges in navigation, visual interpretation, and social interaction. One promising approach to these challenges is employing human guides in VR. However, this approach faces limitations with a lack of availability of humans to serve as guides, or the inability to customize the guidance a user receives from the human guide. We introduce an AI-powered guide to address these limitations. The AI guide features six personas, each offering unique behaviors and appearances to meet diverse user needs, along with visual interpretation and navigation assistance. We aim to use this AI guide in the future to help us understand BLV users’ preferences for guide forms and functionalities.
\end{abstract}




\maketitle
\section{Introduction}
Virtual reality (VR) has been gaining popularity, particularly with the release of affordable headsets such as the Meta Quest Two. As VR gains traction, many of its users have also begun using VR for social interaction. Social VR platforms like VRChat are among the most widely-used VR applications, and saw record-breaking numbers of users over the pandemic \cite{vrchatrecords} while maintaining around 25,000 monthly players as of 2024 \cite{vrchatsteam}. Despite this popularity, these platforms remain largely inaccessible for blind and low vision (BLV) users.

Since the 1990’s, researchers have designed accessible VR environments for BLV people. Some have designed audio or haptic environments specifically for BLV people \cite{lahav2012virtual, lahav2018virtual, sanchez2010metro}, while others added accessibility features to visual environments \cite{trewin2008powerup, andrade2018echo, wedoff2019virtual, blindswordsman, gluck2021racing}. In both cases, the general approach has been to add spatial sounds to objects and actions. For example, a bird flying overhead would generate a “tweet” sound and as the user steps toward it, the sound of footsteps is played. While such features can support BLV users in simple environments, the approach becomes overwhelming in complex and unpredictable ones. Social VR platforms involve a diverse set of environments with multiple interactable objects and users acting in unpredictable ways. Moreover, users speak with one another in social VR environments; audio accessibility features must be minimal to allow BLV users to participate in conversations.

To address the challenges of complex social VR environments, we previously proposed a new approach to VR accessibility using a guide \cite{collins2023guide}. We developed a framework that emulated the sighted guide technique in the physical world. This allowed a BLV user to experience a VR environment with another person serving as their guide. The framework enabled the guide to assist with navigation while the user converses with the guide. We found this approach was promising, but somewhat limited. For example, human guides may not always be available, and some BLV participants felt human assistance hindered their sense of independence. 

To address these limitations, we now propose an AI-powered guide to serve as a “sighted guide” for BLV people in social VR. Since social VR platforms often include various whimsical avatars, we explored guide personas representing options for the guide’s form, behavior, and communication style. In this demonstration, we present six guide personas, describe the guide’s capabilities, design, and implementation, and conclude with plans for future work. This initial implementation of an AI guide presents an additional step forward for VR accessibility. 

\section{\label{Related Work}Related Work}
Researchers have explored accessible VR environments for decades. In early work by Lahav et al. and Sánchez et al., the authors designed audio and haptic virtual environments to help BLV people learn to navigate physical environments such as a metro system \cite{lahav2012virtual, lahav2018virtual, sanchez2010metro}. Other researchers like Trewin et al.\cite{trewin2008powerup} and Andrade et al. \cite{andrade2018echo} added features to visual environments to support navigation and object interaction in VR games \cite{trewin2008powerup, andrade2018echo, wedoff2019virtual, blindswordsman}. These early works involved various VR setups, and similar approaches adding nonvisual feedback to virtual environments have also been incorporated with modern headsets \cite{cosmo, jiang2023beyond, gluck2021racing}.

With the recent rise of social VR, researchers shifted their focus to supporting accessible interpersonal interactions in VR. Several have explored how to make social cues like gestures or proxemics accessible through audio and haptics \cite{dang2023opportunities, balasubramanian2023enable, segal2024socialcueswitch, wieland2022non, ji2022vrbubble, collins2023making}. For instance, Segal et al. developed a toolkit to augment nonverbal behaviors like head nodding and shaking with haptics and audio, making them perceivable to BLV users \cite{segal2024socialcueswitch}. Ji et al.’s VRBubble allowed BLV users to perceive their social distance to avatars via three types of audio feedback that provided information about how far avatars were from them \cite{ji2022vrbubble}. 

We introduced a novel approach to making social VR accessible in our prior work by Collins and Jung et al., where the accessibility solution itself was social, bringing a sighted person to provide navigation and visual interpretation in VR \cite{collins2023guide}. However, our previous paper only explored one guidance scenario with one sighted guide. Considering the complex environments of social VR, the same type of guide may not be useful across every social scenario. We aim to address this in our current work by introducing a guide with six personas, which may help identify which types of guides are preferred in which situations. This persona approach may also develop our understanding of an ideal guide for social VR.

\section{\label{AI Guide Features}AI Guide Features}
Based on our prior work on human guides in VR \cite{collins2023guide}, we developed four features for our AI guide to make it an effective tool for exploring social VR: (1) \textbf{visually interpreting} the environment, (2) \textbf{navigating} the user to specific places or objects, (3) \textbf{adding audio beacons} to objects, and (4) \textbf{changing its form} to suit the user’s needs.

The features work as follows:
\begin{itemize}
    \item \textbf{Visual interpretation}: The guide can converse with the user and answer questions about the environment, both general questions such as “What does this place look like?”, or specific questions like “How far is X object from me?” One of the benefits of using a sighted guide in the physical world is the ability to converse with the guide and ask questions \cite{howtosightedguide}. In this way, BLV people can learn about visual aspects of the environment.
    \item \textbf{Navigation}: The guide can move the user’s avatar by teleporting or walking the user to a specified destination (for example, the user can request,“take me to the fountain” or “take me to the round blue thing”). The user must virtually “grab” the guide to initiate the movement. As the user grabs the guide, the user feels a vibration on the controller. When arriving at the destination, an audio effect is played. Such assisted movement is another core benefit of sighted guidance in the physical world \cite{howtosightedguide}.
    \item \textbf{Audio beacons}: The guide can place an audio beacon on a specified object. Audio beacons have been proposed to help BLV users localize themselves and target objects in the physical world \cite{collins2023guide, soundscape, collins2023making}. They have been found to be an effective way to help BLV orient to objects in their environments for self-guidance \cite{soundscape}. Thus, we added them to allow users to leverage the guide to support independent navigation as well.
    \item \textbf{Changing personas}: The guide can change its persona (form, behavior, and communication style). These guides are meant for use in social VR, which means they could be used in a variety of social situations, from intimate get-togethers with friends or family to professional networking events. As such, we included this feature to let the user change our guide to suit their needs in different social situations (e.g., using a dramatic bird around friends or family versus using a professional white cane around co-workers).
\end{itemize}

Besides these features, we also added audio enhancements to enable BLV people to use the guide. We included unique audio for the user’s actions, such as sounds for teleportation, the user’s footsteps, and to indicate the user turning different directions. We also included footstep sound effects for the guide that change with the guide’s form (e.g., metallic footsteps for a robot persona, flapping wings for a bird persona) to indicate its presence to the user. Finally, we included unique audio that plays as the guide processes the user’s request, and a separate sound that plays before it responds. This provides feedback to the user that they have successfully completed their query while they await the guide’s response.

When a BLV user enters a VR environment, the guide automatically joins them and follows at a specified distance (this differs for each persona, but is generally an arm’s length away). The user can speak to the guide at any time by pressing a button on their controller. The guide can answer questions, add audio beacons, or offer navigation assistance in response. The guide continues to move with the user, unless the user has requested navigation assistance (in this case, the guide will wait for the user to grab it before taking them to their destination, then resume following them). Finally, when the user exits the environment, the guide leaves as well.

The following is an example dialogue between a user and the human persona guide (see section \ref{Personas}), pulled from a final quality test during the development of the guide. We include this dialogue below to demonstrate the kind of output the guide provides:

\textit{User: Hello. Can you tell me what’s going on?}

\textit{Human Guide: Hello! It's splendid to meet you. You're currently standing within a colorful town filled with a variety of shapes, from intriguing buildings to unique vehicles. Right in front of you, almost within arm's reach, is Tall Building in a bright yellow color. Just beyond that is Short Building, a smaller cube but equally vibrant, drenched in green. Not too far from where you are, Sideways Building is resting on its side, creating an elongated yellow contrast with the ground. Move a bit and you'll see Red Car, a cylindrical form waiting for action. In the distance, Landmark stands out, its flattened green oval shape marking a significant spot in our little town. If you'd like to explore any of these or need help finding something specific, just let me know! How can I assist you further on this fine day?}

\textit{User: Uh, can you tell me what the yellow thing in front of me is?}

\textit{Human Guide: The yellow thing in front of you is called Sideways Building. Is there anything else you would like to know or do with Sideways Building?}

\textit{User: Can you take me to Sideways Building?}

\textit{Human Guide: Okay. Grab onto me and I will take you to Sideways Building.} *User presses grip button to grab the guide, both avatars walk to Sideways Building, an arrival sound effect plays.*

\textit{User: Thank you.}

\textit{Human Guide: You are welcome. Let me know if there is anything else I can do.}

\section{\label{Personas}Personas}
Social VR users often experiment with avatar appearance and accessories. In particular, people with disabilities may want to express their disability in different ways depending on context \cite{ zhang2022s, gualano2023invisible}. Since an AI guide is a virtual entity, we can leverage the flexibility of virtuality to explore different guide personas that allow users to customize the presentation of their guide as well as the type of guidance they receive. 

We developed six guide personas derived from prior work on human guides and conversations within our mixed ability team \cite{collins2023guide, zhang2022s, gualano2023invisible}. To design the personas, we brainstormed descriptors that characterized the personalities we wanted each to embody, explained below.

\textbf{Human.} We incorporated a human persona to simulate a sighted guide in the physical world.  BLV users are likely comfortable using human avatars since they are used to talking with humans, making it an effective avatar to introduce the concept of guides in VR. To reduce possible gender bias, we gave our guide an androgynous appearance, with short black hair, a light gray hoodie to conceal their figure, and black pants. This guide followed the user from behind like an assistant, using an androgynous voice that sounded light and friendly. Finally, we used the following descriptors: “warm, friendly, but still professional sighted guide.”

\textbf{Guide Dog.} We based our second guide persona on a guide dog, another type of guide used by BLV people. While the concept of a guide dog is familiar to BLV people, a talking dog may help them imagine other fantastical guide personas. The guide dog avatar also serves as a disability signifier, which has been frequently requested by BLV users for disability representation in social VR \cite{collins2023guide, zhang2022s}. Its avatar was that of a mid-sized dog with a guide dog’s harness and handle, following the user on their left side, and walking so their handle is near the user’s hand. We selected an airy and enthusiastic voice and descriptors that we felt embodied the stereotypical friendly dog: “very friendly, excited companion, who is eager to please who you're talking to.”

\textbf{White Cane.} The third guide persona was based on a common accessibility tool: the white cane. The white cane introduced BLV users to non-living entities as guides, showing that guides could be embodied as tools or accessories. Like the guide dog, the white cane is also a disability signifer noted in prior work \cite{collins2023guide, zhang2022s}. We designed the cane guide to look like a standard mobility cane, with a red stripe in the middle and a large rounded tip, positioned in front of the user. We selected a masculine voice that sounded serious and professional after an informal survey identified this voice as the most monotone among our set of available voices. We used descriptors based on how we imagined a cane would speak, reporting facts quickly and without emotion: “computer-like, succinct assistant, who gives the straight facts.”

\textbf{Robot.} Since the AI guide was a sort of virtual robot, we also represented it as a stereotypical “AI” robot. This avatar also demonstrated a guide that was not possible in the physical world. Our robot guide looked like media depictions of robots: mechanical though humanoid in stature with a broad-shouldered, masculine frame. Being humanoid, the robot followed the user like the “human” guide. We selected a masculine, monotone voice that enunciated its words in a robotic way to match the robot’s appearance. We used descriptors for a stereotypical robot: “formal and assertive assistant, who talks like a robot.”

\textbf{Bird.} Our fifth persona was inspired by our previous study on human guides, where four participants requested bird guides \cite{collins2023guide}, stating they would be more engaging than other forms like the human or the dog. This persona also echoed popular media depictions of birds as guides for blind characters \cite{berto, ayden}. Finally, by including a bird avatar, we could examine if there were any advantages to smaller guides for BLV users with residual vision, since the bird’s size would block less of our users’ field of view. Our bird looked like a phoenix perched on the user’s shoulder, which drew from fantasy, medieval settings that often depict wise birds. We gave the bird a masculine voice with a British accent and a sophisticated style of speech. We used the following descriptors: “wise, old-fashioned, slightly Shakespearean-sounding mentor.” 

\textbf{Invisible.} Our final persona was an invisible guide, following our previous guide study and prior work on representing disability, which found users wish to choose when to represent their disability (or guide) to others \cite{collins2023guide,zhang2022s, gualano2023invisible}. The invisible guide allowed users to keep their disability private, as well as explore VR scenes without the disruption of an avatar following them everywhere. This guide’s invisible form followed the user like the “human” guide. Its voice was feminine and airy–the least distracting voice according to our informal surveys. We used descriptors to embody a secretive manner of speaking: “gentle, soft-spoken assistant who gives very brief statements, as though slipping in words to someone without trying to interrupt what they're doing.” 

\section{\label{AI Integration and Implementation}AI Integration and Implementation}
The AI guide runs on a Meta Quest 2 headset in a virtual environment developed in Unity. The guide’s navigation is handled by Unity’s AI pathfinding components, NavMesh and NavAgents. The guide’s ability to interpret user requests is powered by two APIs from OpenAI: GPT-4V and the Audio API. We use the Speech-to-Text model from the Audio API to translate the user’s questions into a textual query. We then use GPT-4 to process images of the environment alongside the user’s query, and generate the guide’s responses. We engineered our prompt to GPT-4 so that it classifies queries into five categories: (1) holistic descriptions, (2) specific visual questions,(3) requests to go to particular objects, (4) requests to add audio beacons to particular objects, and (5) any other queries (e.g., interpersonal questions such as “How are you doing?”). Finally, we use the Text-to-Speech model from the Audio API to translate the guide’s textual response into speech played for the user. We send the same prompt instructions to GPT-4 each time the user makes a new request, to ensure reliability in the guide’s response. 

To test our system, the sighted researcher developing the prototype created various VR test scenes and repeatedly queried the guide for assistance in these scenes. The test queries were drawn from questions BLV users had asked of the human guide in our prior work \cite{collins2023guide}. The goal of these tests was to ensure there were no technical issues with the system. At weekly intervals throughout the project, the developing researcher would report on the system’s output to the mixed ability team, and make alterations based on feedback from a BLV supervisor of the project. The goal of these alterations was to ensure the system would provide output that was usable, accessible, and understandable to a BLV user. We demonstrated the guide at XR-focused accessibility symposiums where it was used by several BLV users. As a next step, we will conduct formal studies to evaluate the guide.

During our testing, we encountered challenges integrating AI components of our architecture, discussed below.

\subsection{Answering User Questions}
GPT-4 is trained to provide information based on data from the Internet, not to interpret VR scenes, so it initially struggled with many of our requests. For instance, when we sent images of the VR scene from the user’s perspective, GPT-4 described everything as a “screenshot of a video game,” rather than describing the image contents. To address this, we altered our prompt to state explicitly that we sent screenshots of a game. We then instructed GPT-4 to “imagine as though a player in the game asked” the user’s actual query.  This prompt produced descriptions of the screenshots’ contents.

Even after these changes, GPT-4 struggled with questions about the user’s location, since it was only perceiving the environment from the user’s perspective. For example, if a user asked “What’s this yellow stuff in front of me?” or “Where are we in the scene?”, GPT-4 often mistook similar-appearing objects for each other since it could only see one angle of the object, or it assumed the user was always at the start of the virtual environment. We created a second camera in Unity to address location challenges, which captured a bird’s eye view of the scene. We re-contextualized our prompt, telling GPT-4 that “the two photos you are seeing are two views of a video game. One of these photos is the bird's eye view of the entire scene. The other photo is the player's current perspective and what they are currently looking at in the scene.” We took screenshots from both cameras every ten seconds to update the user’s location. With this change, we found that GPT-4 could more accurately interpret where a user was in the virtual scene by figuring out which objects the user was most likely looking at out of those contained in the bird’s eye view. This also allowed the system to recognize that the user had moved around to different objects in the scene, and wasn’t stationary at the scene’s beginning.

Finally, for questions that fell outside guide-specific abilities, we asked GPT-4 to “address the player’s question as best as you can,” allowing it to respond to interpersonal queries.

\subsection{Navigation Requests and Audio Beacons}
Another challenge we encountered was processing action requests, such as a user’s request to navigate to a certain location or place a beacon. The user may not refer to the object by its actual name, or clearly articulate what action they want the guide to perform (e.g., “take me to that red thing over there”). We could not use GPT-4 to process such requests because it could not perform actions. We thus used GPT-4 to identify the name of the target object and action requested, and used Unity to perform the action. 

To enable GPT-4 to identify the target object by name, we first provided it with a list of all the major objects in the environment, along with descriptions of them. We then asked it to tell us whether “it seems” like the user wanted to teleport, walk, or add a sound to the object; we also asked it to provide the name of the object, pulling the name from the list provided. This was added after the guide provided output like, “sideways yellow building, teleport” instead of “Sideways Building, teleport.” We found that the provided descriptions helped GPT-4 identify the object. They can be viewed as “alt text” that a developer must provide to make their environment accessible.

After GPT-4 returned the target object and action in the user’s request, we performed the action in Unity. For navigation, we activated pathfinding algorithms to send the guide’s avatar to the target object. For audio beacons, we created a temporary Audio Source on the target object. One additional behavior we added with navigation was ensuring the guide would wait for the user to grab it before moving. We did this by first deactivating pathfinding scripts that made the guide follow the user, then activating a collider on the guide to make it grabbable. This collider was only activated if GPT-4 determined the user was trying to go somewhere. Once the user grabbed the guide, we called new pathfinding algorithms to take both avatars to the target destination. Upon release, the guide’s collider turns off and it resumes its following algorithm.

\section{\label{Future Work}Future Work}
In the future, we aim to determine preferred personas for an AI guide based on feedback from BLV users. We anticipate users will desire multiple guide personas, which they can use in different social VR contexts. Personalization is important for many users, who may see their guide as an extension of their avatar, or a way to represent (or conceal) their disability. In addition, having guide options to switch between can help the guide adapt to its users’ changing needs, such as needing more interactive guidance in unfamiliar areas or a hands-off approach for familiar worlds. The guide’s appearance is also important to explore, where an invisible guide could allow people to conceal that they are being guided or where visible guides help others understand BLV users’ needs. Finally, it is important to consider users who want to form a connection with their guide, whether that is by giving the guides more “personality” to connect with, or by having the AI guides simply be entertaining, secondary options when human guides are unavailable. We hope to use AI guides to explore these and more questions about guide personas and their impact on the experience of BLV users in social VR.

\begin{acks}
This paper is based on work supported by the National Science Foundation under Grant No. 2212396, and a gift from Meta (Meta Platforms, Inc.). This research was also funded in part by the Office of Naval Research Long Range Broad Agency Announcement (BAA) N00014-23-1-2420. The contents of this paper may not represent the policy of the funders, and no endorsement should be assumed. We thank all the participants for their time.  
\end{acks}

\bibliographystyle{ACM-Reference-Format}
\bibliography{references}

\end{document}